\newif\ifdraft

\draftfalse

\documentclass[journal]{IEEEtran}

\usepackage[dvipsnames]{xcolor}
\usepackage{ulem} 
\usepackage{pdflscape}
\usepackage{graphicx}
\usepackage{hyperref}
\usepackage{verbatim}
\usepackage{mathtools}

\usepackage{subfig} 
\DeclarePairedDelimiter{\ceil}{\lceil}{\rceil}

\newcommand{\gettitle}{Detecting and interpreting myocardial infarction using fully convolutional neural networks}

\graphicspath{{./figures/}}

\hypersetup{
	pdftitle={\gettitle},
	pdfauthor={Nils Strodthoff and Claas Strodthoff},
	pdfkeywords= {electrocardiography} {myocardial infarction} {time series classification} {convolutional neural networks} {interpretability},
	bookmarksopen=false,
	bookmarksnumbered=true
}

\ifdraft

\newcommand{\deletedL}[1]{\textcolor{blue}{\sout{#1}}}
\newcommand{\deletedC}[1]{\textcolor{red}{\sout{#1}}}

\newcommand{\deletedfloatC}[1]{\textcolor{red}{#1}}
\newcommand{\deletedfloatL}[1]{\textcolor{blue}{#1}}
\newcommand{\commentC}[1]{\textcolor{red}{#1}}
\else

\newcommand{\deletedL}[1]{}
\newcommand{\deletedC}[1]{}

\newcommand{\deletedfloatC}[1]{}
\newcommand{\deletedfloatL}[1]{}
\newcommand{\commentC}[1]{}
\fi

\begin{document}

\title{\gettitle}

\author{\thanks{This is the version of the article before peer review or editing, as submitted by an author to Physiological Measurement. IOP Publishing Ltd is not responsible for any errors or omissions in this version of the manuscript or any version derived from it. The Version of Record is available online at \href{https://doi.org/10.1088/1361-6579/aaf34d}{10.1088/1361-6579/aaf34d}.}Nils~Strodthoff\thanks{Nils Strodthoff is with Fraunhofer Heinrich Hertz Institute, 10587 Berlin, Germany, e-mail: nils.strodthoff@hhi.fraunhofer.de.}\,\&\,
Claas~Strodthoff\thanks{Claas Strodthoff is with Department of Anesthesiology and Intensive Care Medicine, University Medical Center Schleswig-Holstein, Campus Kiel, Kiel, Germany, e-mail: claas.strodthoff@uksh.de.}}
\maketitle

\begin{abstract} 
\textit{Objective:} We aim to provide an algorithm for the detection of myocardial infarction that operates directly on ECG data without any preprocessing and to investigate its decision criteria.\\
\textit{Approach:} We train an ensemble of fully convolutional neural networks on the PTB ECG dataset and apply state-of-the-art attribution methods.\\
\textit{Main results:} Our classifier reaches 93.3\% sensitivity and 89.7\% specificity evaluated using 10-fold cross-validation with sampling based on patients. The presented method outperforms state-of-the-art approaches and reaches the performance level of human cardiologists for detection of myocardial infarction. We are able to discriminate channel-specific regions that contribute most significantly to the neural network's decision. Interestingly, the network's decision is influenced by signs also recognized by human cardiologists as indicative of myocardial infarction.\\
\textit{Significance:} 
Our results demonstrate the high prospects of algorithmic ECG analysis for future clinical applications considering both its quantitative performance as well as the 
possibility of assessing decision criteria on a per-example basis, which enhances the comprehensibility of the approach.
\end{abstract}

\section{Introduction}
Ischaemic heart diseases are the leading cause of death in
Europe. The most prominent entity of this group is acute myocardial infarction (MI),
where the blood supply to parts of the heart muscle is permanently interrupted due
to an occluded coronary artery.
Early detection is crucial for the effective treatment of acute myocardial
infarction with percutaneous coronary intervention (PCI) or coronary artery bypass
surgery. Myocardial infarction is usually diagnosed with the help of clinical findings,
laboratory results, and electrocardiography. ECGs are produced by recording
electrical potentials of defined positions of the body surface over time,
representing the electric activity of the heart. Deviations from the usual shape
of the ECG curves can be indicative of myocardial infarction as well as many
other cardiac and non-cardiac conditions. ECGs are a popular diagnostic tool as
they are non-invasive and inexpensive to produce but provide a high diagnostic value.

Clinically, cases of myocardial infarction fall into one of two categories, ST-elevation
myocardial infarction (STEMI) and non-ST-elevation myocardial infarction (NSTEMI),
depending on whether or not the ECG exhibits a specific ECG-sign called ST elevation. The
former can and should be treated as soon as possible with PCI, whereas the NSTEMI diagnosis
has to be confirmed with time-costly laboratory tests before specific treatment can be initiated \cite{pmid26727671}.
Since waiting for these results can delay effective treatment by hours, a more detailed
analysis of the ECG could speed up this process significantly.

Failure to identify high-risk ECG findings in the emergency department is common
and is of grave consequences \cite{Masoudi2006}. To increase accuracy, speed and
economic efficiency, different algorithms have been proposed to automatically
detect myocardial infarction in recorded ECGs. Algorithms with adequate
performance would offer significant advantages: Firstly, they could be applied
by untrained personnel in situations where no cardiologist is available. Secondly, once
set up they would be highly reliable and inexpensive. Thirdly, they could be tuned to specific
decision boundaries, for example to high sensitivity (low specificity) for
screening purposes.

Common ECG classification algorithms usually mimic the approach a human
physician would take: First preprocessing steps include the correction of baseline
deviations, noise reduction and the segmentation of single heartbeats. 
In the next step, hand-engineered features such as predefined or
automatically detected time intervals and voltage values are extracted from the
preprocessed signal. Finally, the classification is carried out with a variety
of common classifiers such as simple cutoff values, support vector machines or
neural networks. Preprocessing and feature extraction are non-trivial steps with
technical and methodical problems, especially with unusual heart rhythms or
corrupted data, resulting in a high risk of information loss. This urges for a
more unified and less biased algorithmic approach to this problem.

Deep neural networks \cite{lecun2015deep,Goodfellow-et-al-2016} and in particular
convolutional neural networks have been the driving force behind the tremendous
advances in computer vision \cite{krizhevsky2012imagenet,szegedy2015going,2015arXiv151203385H,erhan2014scalable,ren2015faster,long_fully_2015}
in recent years. Consequently, related methods have also been applied to the problem of time series
classification in general and ECG classification tasks specifically. Even though
we focus exclusively on ECG classification in this work,
we stress that the methodology put forward here can be applied to generic
time series classification problems in particular to those that satisfy the following conditions:
Data (a) with multiple aligned channels (b) that arises as a continuous sequence, i.e.\ with no start/end points in the sequence), and 
that exibits a degree of periodicity, and (c) that is fed to the algorithm in unprocessed/unsegmented form.
These three criteria define a subclass within general time series classification
problems that is important for many real-world problems. In particular, these criteria include
raw sensor data from medical monitoring such as ECG or EEG.

The main contributions presented in this paper are the following:
\begin{enumerate}
\item We put forward a fully convolutional neural network for myocardial infarction detection on the PTB
dataset \cite{PTB,PhysioBank} focusing on the clinically most relevant case of
12 leads. It outperforms state-of-the-art literature approaches \cite{sharma2018inferior,Reasat2017}
and reaches the performance level of human cardiologists reported in an earlier
comparative study \cite{willems1991diagnostic}.

\item We study in detail the classification performance on subdiagnoses and investigate
channel selection and its clinical implications.

\item We apply state-of-the-art attribution methods to investigate
the patterns underlying the network's decision and draw parallels to cardiologists'
rules for identifying myocardial infarction.
\end{enumerate}

\section{Related works}
Concerning time series classification in general, we focus on time series
classification  using deep neural networks and do not discuss traditional
methods in detail, see e.g.\ \cite{Bagnall2016} for a recent review. H\"usken
and Stagge \cite{Husken2003} use recurrent neural networks for time series
classification. Wang et al \cite{Wang2016} use different mainly convolutional
networks for time series classification and achieve state-of-the-art results in
comparison to traditional methods applied to the UCR Time Series Classification
Archive datasets \cite{UCRArchive}. Cui et al \cite{Cui2016} use a sliding
window approach similar to the one applied in this work and feed differently
downsampled series into a multi-scale convolutional neural network also reaching
state-of-the-art results on UCR datasets. Also, recurrent neural networks have
been successfully applied to time series classification problems in the clinical
context \cite{DBLP:journals/corr/ChoiBS15}. More recent works include attention
\cite{2017arXiv171103905S} and more elaborate combinations of convolutional and
recurrent architectures \cite{DBLP:journals/corr/abs-1709-05206,2018arXiv180104503K}.
For a more detailed account on deep learning methods for time series classification, 
we refer the reader to the recent review \cite{2018arXiv180904356I}.

Concerning time series classification on ECG signals, the two main areas of work are the detection of either arrhythmia or infarction.
It is beyond the scope of this work to review the rich body of literature on classification
of ECG signals using algorithmic approaches, in particular those involving neural networks, see e.g\ \cite{Maglaveras1998ECGPR} for a classic review. 
While the literature on myocardial infarction is covered in detail below, we want to briefly mention some more recent works \cite{rajpurkar_cardiologist-level_2017,ACHARYA201781,ACHARYA2018952,xia2018detecting} in the broad
field of arrhythmia detection. For further references we refer the reader to the recent reviews \cite{Luz2016ECGbasedHC,HAGIWARA201899}. At this point we also want to highlight \cite{rajpurkar_cardiologist-level_2017}, where the authors trained a convolutional 
neural network on an exceptionally large custom dataset reaching human-level performance in arrhythmia detection. 

More specifically, turning to myocardial infarction detection in ECG recordings,
many proposed algorithms rely on classical machine learning methods for
classification after initial preprocessing and feature extraction
\cite{Sun2012,Arif2012,Lee2013,safdarian2014new,kojuri2015prediction,liu2015novel,Negandhi2016}.
Particular mentioning deserve \cite{Sharma2015multiscale,sharma2018inferior} who operate on Wavelet-transformed signals.
Whereas the above works used neural networks at most as a classifier on top of
previously extracted features
\cite{Arif2010,safdarian2014new,Kora2015,kojuri2015prediction}, there are
works that apply neural networks directly as feature  extractors to
beat-level separated ECG signals \cite{acharya2017application, Kachuee2018}.
These have to be distinguished from approaches as the one considered in this
work where deep neural networks are applied to the raw ECG with at most minor
preprocessing steps. In this direction Zheng et al \cite{Zheng2014}, present an
approach based on convolutional neural networks for multichannel time series
classification similar to ours but applied it to ECGs in the context of
congestive heart failure classification. Also \cite{ACHARYA201781, ACHARYA201762, ACHARYA2018952} use a related approach 
for arrhythmia/coronary artery disease detection in a single-channel setting. The most recent work
on the myocardial infarction detection using deep neural networks
\cite{Reasat2017} also uses convolutional architectures applied to
three channel input data. A quantitative comparison to their results is
presented in Sec.~\ref{sec:results}. Similar performance was reported in
\cite{2018arXiv180206458R} who used LSTMs on augmented channel data obtained from a
generative model.

\section{Dataset and medical background} \label{sec:data}

The best-known collection of standard datasets for time series classification is
provided by the UCR Time Series Classification Archive \cite{UCRArchive}.
Although many benchmarks are available for the contained datasets
\cite{Bagnall2016}, we intentionally decided in favor of a different dataset, as
the UCR datasets contain only comparably short and not necessarily periodic
sequences and are almost exclusively single-channel data. The same applies to
various benchmarks datasets \cite{UCRArchive, DBLP:journals/corr/PeiDTM15,
DBLP:journals/corr/abs-1711-11343} considered for example in
\cite{2018arXiv180104503K}, which do not match the criteria put forward in the introduction. In
particular the requirement of continuous data with no predefined start and end
points that show a certain degree of periodicity is rarely found in existing
datasets, especially not in combination with the other two requirements from above.

We advocate in-depth studies of more complex datasets that are more
representative for real-world situations and therefore concentrate our study on
ECG data provided by the PTB Diagnostic ECG Database \cite{PTB,PhysioBank}. It
is one of the few freely available datasets that meet the conditions from
above. The dataset comprises 549 records from 290 subjects. For this study we only aim
to discriminate between healthy control and myocardial infarction. Therefore, we
only take into account records classified as either of these two diagnosis
classes. We excluded 22 ECGs from 21 patients with unknown localization and
infarction status from our analysis.

For some patients classified as myocardial infarction the dataset includes
multiple records of highly variable age and in some cases even ECGs recorded
after the medical intervention. The most conservative choice would be to exclude
all myocardial infarction ECGs after the intervention and within a preferably
short threshold after the infarction. Such a dataset would be most
representative for the detection of acute myocardial infarction in a clinical
context, would, however, seriously reduce the already small dataset. As a
compromise, we decided to keep all healthy records but just the first ECG from
patients with myocardial infarction. Note that a selection based on ECG age is
not applicable here as the full metadata is not provided for all records. For
the ECGs where the full metadata is provided this selection leads to a median
(interquartile range) of the infarction age of 2.0 (4.5) days with 14\% of them
taken after intervention. On the contrary including all infarction ECGs would
result in a median of $8.0 (13.8)$ days, $37\%$ of which were taken after
intervention. These figures render the second, most commonly employed, selection
questionable for an acute infarction detection problem.
In summary, our selection leaves us with a dataset of 127 records classified as
myocardial infarction and 80 records (from 52 patients) classified as healthy
control. Demographical and statistical information on the dataset using the selection criteria from above
is compiled in Tab.~\ref{tab:demographics}.

\begin{table}[ht]
    \centering
    \begin{tabular}{l||l|l||l}
    quantity & MI & HC & all\\
    \hline
    \hline
    \# patients & 127 & 52 & 179 \\
    sex: male/female &92/35 & 39/13&131/48\\
    age: median(iqr)/nans & 61.0(16.0)/0 & 38.4(24.0)/6 & 57.0(19.0)/6 \\
    \hline
    MI: untreated/treated/nans\!\! & 81/18/28 &-& -\\
    MI age: median(iqr)/nans\!\! & 2.0(4.5)/4 &-& -\\
    \end{tabular}
    \caption[]{Demographical/statistical information on the selected records from the PTB dataset. Abbreviations: MI: myocardial infarction HC: healthy control iqr:interquartile range nans: records with no information provided \commentC{(this table was newly introduced)}}
    \label{tab:demographics}
\end{table}

\begin{table}[h]
\centering
\begin{tabular}{l||lll}
subdiagnosis/localization & \# patients &\# samples (selected)\\
\hline
\hline
    anterior & 17 & 47 (17) \\
    antero-septal &27& 77 (27) \\
    antero-septo-lateral &1& 2 (1) \\
    antero-lateral &16& 43 (16) \\
    lateral &1& 3 (1) \\
\hline
    $\Sigma$ aMI &62& 172 (62)\\
\hline
\hline
    inferior &30& 89 (30) \\
    infero-posterior &1& 1 (1) \\
    infero-postero-lateral &8& 19 (19) \\
    infero-lateral &23& 56 (23) \\
    posterior &1& 4 (1) \\
    postero-lateral &2& 5 (2) \\
\hline
    $\Sigma$ iMI &65& 174 (65)\\
\hline
\hline
\hline
    Healthy control &52& 80 (80) \\
    $\Sigma$ MI &127& 346 (127)\\
    $\Sigma$ all &179& 426 (207)\\
\end{tabular}

\caption[]{Infarction localization in the PTB ECG Database.}
\label{tab:ptbsub}
\end{table}

For the case of myocardial infarction the dataset distinguishes different
subdiagnoses corresponding to the localization of the infarction, see
Tab.~\ref{tab:ptbsub}, with smooth transitions between certain subclasses. It is
therefore not reasonable to expect to be able to train a classifier that is able
to distinguish records into all these subclasses based on the rather small
number of records in certain cases. We therefore decided to distinguish just two
classes that we colloquially designate as anterior myocardial infarction (aMI)
and inferior myocardial infarction (iMI), see Tab.~\ref{tab:ptbsub} for a
detailed breakdown. This grouping models the most common anatomical variant of
myocardial vascular supply with the left coronary artery supplying the regions
noted in the aMI group and the right coronary artery supplying those in the iMI
group \cite{Moore}. If not noted otherwise we only use the subdiagnoses
information for stratified sampling of records into cross-validation folds and
just discriminate between healthy control and myocardial infarction. In
Sec.~\ref{sec:results_subdiagnoses} we specifically investigate the impact of
the above subdiagnoses on the classification performance. The fact that the
inferior and anterior myocardial infarction can be distinguished rather well
represents a further a posteriori justification for our assignment.

The PTB Database provides 15 simultaneously measured channels for each record: six
limb leads (Einthoven: I, II, III, and Goldberger: aVR, aVL, aVF), six precordial
leads (Wilson: V1, V2, V3, V4, V5, V6), and the three Frank leads
(vx, vy, vz). As the six limb leads are linear combinations of just two measured
voltages (e.g.\ I and II) we discard all but two limb leads. Frank leads are
rarely used in the clinical context. Consequently, in our analysis we only take
into account eight leads that are conventionally available in clinical applications
and non-redundant (I, II, V1, V2, V3, V4, V5, V6). This is done
in spite of the fact that using the full although clinically less relevant set
of channels can lead to an even higher classification performance, see the analysis
in Sec.~\ref{sec:results_medical} where the lead selection is discussed in detail.

\section{Classifying ECG using deep neural networks}
\subsection{Algorithmic procedure}
\label{sec:algorithmic}

As discussed in the previous section, time series classification in a realistic
setting has to be able to cope with time series that are so large that they
cannot be used as input to a single neural network or that cannot be downsampled
to reach this state without losing large amounts of information. At this point two
different procedures are conceivable: Either one uses attentional models that
allow to focus on regions of interest, see e.g.\
\cite{DBLP:journals/corr/abs-1709-05206,2018arXiv180104503K}, or one extracts
random subsequences from the original time series. For reasons of simplicity
and with real-time on-site analysis in mind we
explore only the latter possibility, which is only applicable for
signals that exhibit a certain degree of periodicity. The assumption underlying
this approach is that the characteristics leading to a certain classification
are present in every random subsequence. We stress at this point that this
procedure does not rely on the identification of beginning and endpoints of
certain patterns in the window \cite{Hu2013}. This approach can be justified a
posteriori with the reasonable accuracies and specificities it achieves.
Furthermore, from a medical point of view it is
reasonable to assume that ECG characteristics do not change drastically within
the time frame of any single recording.

The procedure leaves two hyperparameters: the choice of the window size and an
optional downsampling rate to reduce the temporal input dimension
for the neural network. As the dataset is not large enough for extensive
hyperparameter optimizations we decided to work with a fixed window size of 4
seconds downsampled to an input size of 192 pixels for each sequence. The window
size is sufficiently large to capture at least three heartbeats at normal heart
rates.

As discussed in Sec.~\ref{sec:data}, if we consider a binary classification problem
we are dealing with an imbalanced dataset with 80 healthy records
in comparison to 127 records diagnosed with myocardial infarction.
Several approaches have been discussed in the literature to best deal with
imbalance \cite{2015arXiv150501658B,2017arXiv171005381B}. Here we follow the
general recommendations and oversample the minority class of healthy patients by
2:1.

We refrain from using accuracy as target metric as it depends on the ratio of healthy and
infarction ECGs under consideration. As sensitivity and specificity are the most common metrics
in the medical context, we choose Youden's J-statistic as target metric for model
selection which is determined by the sum of both quantities i.e.\
\begin{align}
    J&=\text{sensitivity}\,+\,\text{specificity}\,-1\nonumber\\[2ex]
    &=\frac{TP}{TP+FN}+\frac{TN}{TN+FP}-1\,,
    \label{eq1}
\end{align}
where $TP/FN/FP$ denote true positive/false negative/false positive classification results.
Other frequently considered observables in this context include $F_1$ or $F_2$ scores
that are defined as combinations of positive predictive value (precision) and
sensitivity (recall).

Finally, to obtain the best possible estimate of the test set sensitivity and
specificity using the given data, we perform 10-fold cross-validation on the
dataset. Its size is comparably small and there are still considerable
fluctuations of the final result statistics, even considering the data
augmentation via random window selection. These result statistics do not necessarily reflect the
variance of the estimator under consideration when applied to unseen data
\cite{isaksson2008cross} and it is not possible to infer variance information
from cross-validation scores by simple means \cite{bengio2004no}. The given
dataset is not large enough to allow a train-validation-test split with
reasonable respective sample sizes. Following \cite{SHAIKHINA201751}, we
circumvent this problem by reporting ensemble scores corresponding to models
with different random initializations without performing any form of
hyperparameter tuning or model selection using test set data. Compared to single
initializations the ensemble score gives a more reliable estimate of the
model's generalization performance on unseen data. For calculating the ensemble
score we combine five identical models and report the ensemble score formed by
averaging the predicted scores after the softmax layer
\cite{2017arXiv170401664J}.

\subsection{Investigated architectures}
We investigate both convolutional neural networks as well as recurrent neural network
architectures. While recurrent neural networks seem to be the most obvious choice
for time series data, see e.g.\ \cite{Husken2003}, convolutional architectures have
been applied for similar tasks in early days, see e.g.\ \cite{waibel_phoneme_1989} for
applications in phoneme recognition.

We study different variants of convolutional neural networks inspired by several
successful architectures applied in the image domain such as
fully convolutional networks \cite{long_fully_2015} and resnets
\cite{2015arXiv151203385H,2016arXiv160305027H,2016arXiv160507146Z}, see App.~\ref{app:architectures}
for details. In
addition to architectures that are applied directly to the (downsampled)
time series data, we also investigate the effect of incorporating frequency-domain
input data obtained by applying a Fourier transform to the original time-domain data.
We stress again that our approach operates directly on the (downsampled) input data without 
any preprocessing steps.

For comparison we also consider recurrent neural networks, namely LSTM
\cite{hochreiter_long_1997} cells. We investigate two variants: In the
first approach we feed the last LSTM output into a fully connected layer. In the second case we
additionally apply a time-distributed dense layer i.e.\ a layer with shared weights across all time steps to
train the network in addition on a time-series classification task where we
adjusted both loss functions to reach similar values. Similar to
\cite{Husken2003} we investigate in this way if the time series predication task
improves the classification accuracy.

\section{Results}
\label{sec:results}
\subsection{Network architectures}
\label{sec:results_architectures}

In Tab.~\ref{tab:results_architectures} we compare the architectures described
in the previous section based on 12-lead data. The comparison is based on cross-validated $J-$statistics
without implying statistical significance of our findings. 
In the light of very small number of 20 or even fewer patients in the respective
test sets, we do not report confidence intervals or similar variance measures as these would be
mainly driven by the fluctuations due to the small size of the test set, see also \cite{isaksson2008cross,bengio2004no, repeatedCV} and the 
related discussion in Sec.~\ref{sec:algorithmic}.

The fully convolutional
architecture and the resnet achieve similar performance applied to time-domain
data. In contradistinction to an earlier investigation \cite{Wang2016} that favored the
fully convolutional architecture, a ranking of the two convolutional
architectures is not possible on the given data. Interestingly, the convolutional
architectures perform better applied to raw time-domain data than applied to
frequency-domain data. In this context it might be instructive to investigate
also other transformations of the input data such as Wavelet transformations as
considered in \cite{Sharma2015multiscale,sharma2018inferior}.

Both convolutional architectures show a better score than recurrent architectures. This
can probably be attributed to the fact that we report just results with standard
LSTMs and do not investigate more advanced mechanisms such as most notably an
attention mechanism, see e.g.\
\cite{2017arXiv171103905S,DBLP:journals/corr/abs-1709-05206,2018arXiv180104503K} for recent
developments in this direction. Training the recurrent
neural network jointly on a classification task as well as on a time series
prediction task, see also the description in App.~\ref{app:recurrent}, did not lead to an
improved score, whereas a significant increase was reported by \cite{Husken2003},
which might be related to the small size of the dataset.

\begin{table}[ht]
\centering
\begin{tabular}{l||c||c|c|c}
Model & J-Stat & sens. & spec. & prec. \\
\hline
\hline
\textbf{Fully Convolutional} & \textbf{0.827} & 0.933 & 0.897 & 0.936  \\
\textbf{Resnet} & \textbf{0.828} & 0.925 & 0.903 & 0.940  \\
\hline
\textbf{Fully Convolutional(freq)} & \textbf{0.763} & 0.902 & 0.860 & 0.913  \\
Resnet(freq) & 0.656 & 0.870 & 0.786 & 0.869  \\
\hline
LSTM mode (final output) & 0.743 & 0.910 & 0.833 & 0.899  \\
LSTM (final output + pred.) & 0.742 & 0.914 & 0.828 & 0.897  \\
\end{tabular}
    \caption{Classification results for different network architectures on 12-lead data. Abbreviations: sens.: sensitivity=recall; spec.: specificity; prec.: precision=positive predictive value}
\label{tab:results_architectures}
\end{table}

In the following sections we analyze particular aspects of the classification results in more detail.
All subsequent investigations are carried out using the fully convolutional architecture, which
achieved the same performance as the best performing resnet architecture with a comparably much simpler
architecture. If not noted otherwise we use the default setup of 12-lead data.

\subsection{MI localization, benchmarks, and channel selection} \label{sec:results_subdiagnoses}
\subsubsection{MI localization and training procedure}
As described in Sec.~\ref{sec:data} we distinguish the aggregated subdiagnosis
classes aMI and iMI. Here we examine the classification
performance of a model that distinguishes these subclasses rather than training just on
a common superclass myocardial infarction. We can investigate a number of different
combination of either training/evaluating with or without subdiagnoses as shown in Tab.~\ref{tab:results_subdiagnoses}

\begin{table}[!ht]
\centering
\begin{tabular}{l||c||c|c|c}
Data & J-Stat & sens. & spec. & prec.  \\
\hline
\hline
cardiologists aMI \cite{willems1991diagnostic} & 0.857 & 0.874 & 0.983 & - \\
cardiologists iMI \cite{willems1991diagnostic} & 0.738 & 0.749 & 0.989 & - \\
\hline
train MI eval MI & 0.827 & 0.933 & 0.897 & 0.936  \\ 
\textbf{train MI eval aMI} & \textbf{0.877} & 0.980 & 0.897 & 0.884 \\
\textbf{train MI eval iMI} & \textbf{0.789} & 0.894 & 0.896 & 0.879 \\
\hline
train aMI eval aMI & 0.880 & 0.919 & 0.961 & 0.950 \\
train iMI eval iMI & 0.689 & 0.810 & 0.879 & 0.849 \\
\hline
\hline
train aMI+iMI eval MI & 0.788 & 0.912 & 0.876 & 0.947 \\
train aMI+iMI eval aMI & 0.846 & 0.966 & 0.881 & 0.906 \\
train aMI+iMI eval iMI & 0.741 & 0.861 & 0.879 & 0.897 \\
\end{tabular}

\caption{Classification performance on subdiagnoses (with fully convolutional
architecture) on 12-lead data. Abbreviations as in
    Tab.~\ref{tab:results_architectures}. train MI (train aMI+iMI) refers to training disregarding (incorporating)
subdiagnoses and train aMI/iMI to training using just a particular subdiagnosis; analogously for eval.
 }
\label{tab:results_subdiagnoses}
\end{table}

Both for models trained on unspecific infarction and for models trained
using subdiagnosis labels, the performance on the inferior myocardial infarction
classification task turns out to be worse than the score achieved for
anterior myocardial infarction. The most probable reason for this is that
anterior myocardial infarctions show typical signs in most of the Wilson leads
because of the proximity of the anterior myocard to the anterior chest wall. For
the more difficult task of iMI classification, the model seems to profit
from general myocardial infarction data during training, as a model trained on
generic MI achieves a higher score on aMI classification than a model trained
specifically on aMI classification only. The converse is true for the simpler
task of aMI classification.

Interestingly, the model trained without subdiagnoses reaches a slightly higher
score both for unspecific myocardial infarction classification as well as for
classification on subdiagnoses aMI/iMI only, which might just be an effect of an
insufficient amount of training data. In any case, we restricted the rest of our
investigations on the model trained disregarding subdiagnoses.

\begin{figure}[ht]
	\centering
	\includegraphics[width=.8\columnwidth]{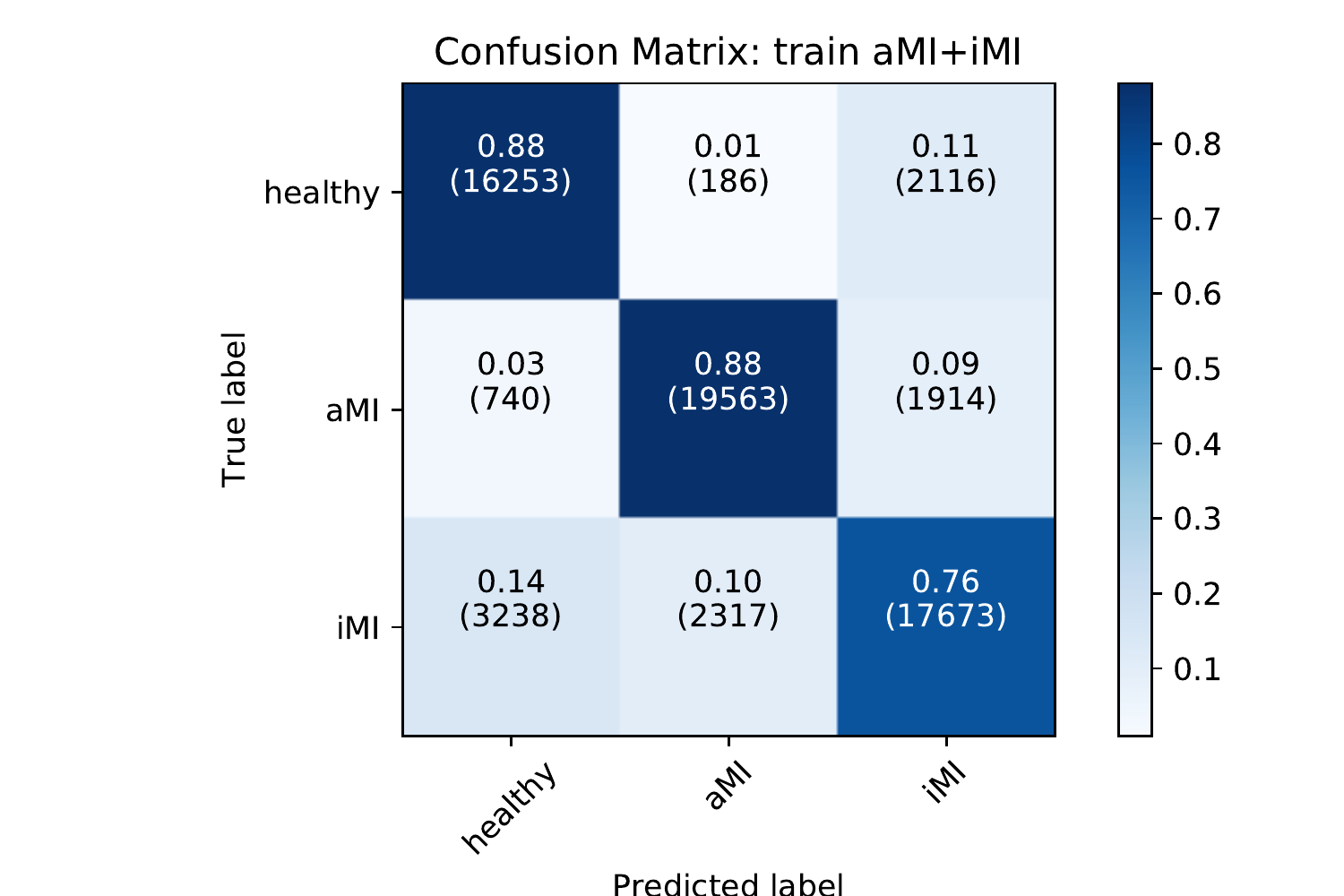}
    \caption{Confusion matrix for model trained on subdiagnoses aMI and iMI.}
    \label{fig:confusion012}
\end{figure}
In Fig.~\ref{fig:confusion012} we show the confusion
matrix for the model that is trained and evaluated on the subdiagnoses aMI and
iMI in addition to healthy control. The confusion matrix underlines the fact
that the model is able to discriminate between the aggregated subdiagnoses whose
assignments were motivated by medical arguments, see Sec.~\ref{sec:data}, and
represents an a posteriori justification for this choice.

The reported score for models evaluated on subdiagnoses allows a comparison to
human performance on this task. We base this comparison on a study
\cite{willems1991diagnostic} that assessed the human classification performance
for different diagnosis classes based on a panel of eight cardiologists. Here we
only report the combined result and refer to the original study for individual
results. The most appropriate comparison is the model trained on general MI and
evaluated on subdiagnoses aMI and iMI. However, it turns out that irrespective
of the training procedure the algorithm achieves a slightly higher score on aMI
classification and a considerably higher score on iMI classification and we see
it therefore justified to claim at least human-level performance on the given classification task.
We restrain from drawing further conclusions from this comparison
as it depends on the precise performance metric under consideration, the fact
that not the same datasets were used in both studies and differences in
subdiagnosis assignments. The claim of superhuman performance on this task would
certainly require more thorough investigations in the future.

\subsubsection{Comparison to literature approaches}
Our data and channel selection strategy, see Sec.~\ref{sec:data}, was carefully
chosen to reflect the requirements of a clinical application as closely as
possible. In addition, considering the comparably small size of the dataset, a
careful cross-validation strategy is of utmost importance, see
Sec.~\ref{sec:algorithmic}. Unfortunately, most literature results do not
report cross-validated scores or introduce data leakage in their cross-validation
procedures. This can happen for example by sampling from beat-level segmented
signals or most commonly by sampling based on ECGs rather than patients, see
\cite{sharma2018inferior} for a detailed discussion. Both
cases lead to unrealistically good performance estimates as the classification
algorithm can in some form adapt to structures in the same ECG or structures in
a different ECG from the same patient during the training phase.

We refrain from presenting results for the latter setups as they do not allow to
disentangle a model's  classification performance from its ability to reproduce
already known patterns. Therefore, we only include a comparison to the most
recent works \cite{sharma2018inferior, Reasat2017} that are to our knowledge the
only works where a cross-validated score with sampling on patient level, in the
literature also termed subject-oriented approach \cite{sharma2018inferior}, is
reported. To ensure comparability with \cite{sharma2018inferior,Reasat2017} we
replicate their setup as closely as possible and modify our data selection to
include limb leads and in addition to healthy records only all genuine inferior
myocardial infarction ECGs. In this case our approach shows not only superior
performance compared to literature results, see Tab.~\ref{tab:results_benchmark}, but does unlike their algorithms 
operate directly on the input data and does not require preprocessing 
with appropriate input filters.

\begin{table}[ht]
\centering
\begin{tabular}{l||c||c|c|c}
Benchmark & J-Stat & sens. & spec. & prec.  \\
\hline
\hline
Wavelet transform + SVM \cite{sharma2018inferior} & 0.583 & 0.790 & 0.793 & 0.803 \\
CNN \cite{Reasat2017} & 0.694 & 0.853 & 0.841 & -\\
\textbf{limb leads + inferior MI$^*$}  & \textbf{0.773} & 0.874 & 0.900 & 0.932\\
\end{tabular}
\caption{Comparison to literature results (with fully convolutional architecture). Abbreviations as in Tab.~\ref{tab:results_architectures}. Results from this work marked by asterisk.}
\label{tab:results_benchmark}
\end{table}

We replicated the above setup to demonstrate the competitiveness of our
approach, but for a number of reasons we are convinced that the scores presented
in Tabs.~\ref{tab:results_architectures} and \ref{tab:results_subdiagnoses} are
the more suitable benchmark results: Firstly, from a clinical point of view
12-lead ECGs are the default choice and the algorithm should be fed with the
full set of 8 non-redundant channels. Secondly, the restriction to include only
the first infarction ECG per patient is arguably more suited for the application
of the clinically most relevant problem of classifying acute myocardial
infarctions, see the discussion in Sec.~\ref{sec:data}. Finally, from a machine
learning perspective it is beneficial to include all subdiagnoses for training
allowing to adapt to general patterns in infarction ECGs and to only evaluate
the trained classifier on a particular subdiagnosis of interest. For the case of
aMI this procedure leads to an improved score, see the discussion of
Tab.~\ref{tab:results_subdiagnoses}.

\subsubsection{Channel selection}
\label{sec:results_medical}

By including different combinations of leads one can estimate the relative amount of information that these channels
contribute to the classification decision, see Tab.~\ref{tab:results_channels}.

\begin{table}[ht]
\centering
\begin{tabular}{l||c||c|c|c}
channels & J-Stat & sens. & spec. & prec.  \\
\hline
\hline
all leads & 0.878 & 0.941 & 0.937 & 0.961  \\
\hline
12 leads (default) & 0.827 & 0.933 & 0.897 & 0.936  \\
\hline
Frank leads only & 0.803 & 0.930 & 0.873 & 0.923  \\
\hline
limb leads only & 0.811 & 0.912 & 0.899 & 0.937  \\
\hline
I only & 0.703 & 0.875 & 0.828 & 0.893  \\
II only & 0.695 & 0.907 & 0.787 & 0.874  \\
III only & 0.590 & 0.855 & 0.735 & 0.841  \\
\end{tabular}
    \caption{Channel dependence of the classification performance (with fully convolutional architecture). Abbreviations as in Tab.~\ref{tab:results_architectures}.}
\label{tab:results_channels}
\end{table}

Starting with single-lead classification results, out of leads I, II and III, lead III offers the least
amount of information, possibly because its direction coincides worst with the
usual electrical axis of the heart. The classification result using Frank leads achieves
a score that is slightly worse than the result using limb-leads only. A further performance
increase is observed when complementing the limb leads with the Wilson leads towards the standard 12-lead setup.
The overall best result is achieved using all channels, which does, however, not correspond to
the clinically relevant situation, where conventionally only 12 leads are available.

\subsection{Interpretability} \label{sec:results_interpretability}
A general challenge remains the topic of interpretability of machine learning algorithms
and in particular deep learning approaches that is especially important for
applications in medicine \cite{doi:10.1001/jama.2017.7797}. In the area of deep
learning, there has been a lot of progress in this direction \cite{2013arXiv1312.6034S, 2017arXiv170301365S, MONTAVON20181}. So far
most applications covered computer vision whereas time series data in particular did only
receive scarce attention. Interpretability methods have been applied to time series data in \cite{Wang2016}
and ECG data in particular in \cite{2018arXiv180205998T}. A different approach towards interpretability in
time series was put forward in \cite{2018arXiv180202952S}.

As an exploratory study for the application of interpretability methods to time series
data we investigate the application of attribution methods to the trained classification model.
This allows investigating on a qualitative level if the machine
learning algorithm uses similar features as human cardiologists. Our
implementation makes use of the DeepExplain framework put forward in
\cite{DBLP:journals/corr/abs-1711-06104}. For neural networks with only ReLU
activation functions it can be shown \cite{DBLP:journals/corr/abs-1711-06104}
that attention maps from `$\text{gradient}\times \text{input}$'
\cite{DBLP:journals/corr/ShrikumarGSK16} coincide with attributions obtained
via the $\epsilon$-rule in LRP \cite{bach-plos15}. Even though we are using ELU
activation functions the attribution maps show only minor quantitative
differences. The same applies to the comparison to integrated gradients
\cite{2017arXiv170301365S}. For definiteness, we focus our discussion on
`$\text{gradient}\times \text{input}$'. Different from computer vision, where
conventionally attributions of all three color channels are summed up, we keep
different attributions for every channel to be able to focus on channel-specific
effects. We use a common normalization of all channels to be able to compare
attributions across channels.

We stress that attributions are inherent properties of the underlying models and
can therefore differ already for models with different random initializations in
an otherwise identical setup. If we aim to use it to identify typical
indicators for a classification decision as a guide for clinicians a more
elaborate study is required. For simplicity, in this exploratory study we focus on a single model rather than the model ensemble. By visual inspection we
identified the most typical attribution pattern for myocardial infarction among
examples in the batch that occurred shortly after the infarction. Prototypical
outcomes of this analysis are presented below.

\begin{figure}[ht]
	\centering
	\includegraphics[width=\columnwidth]{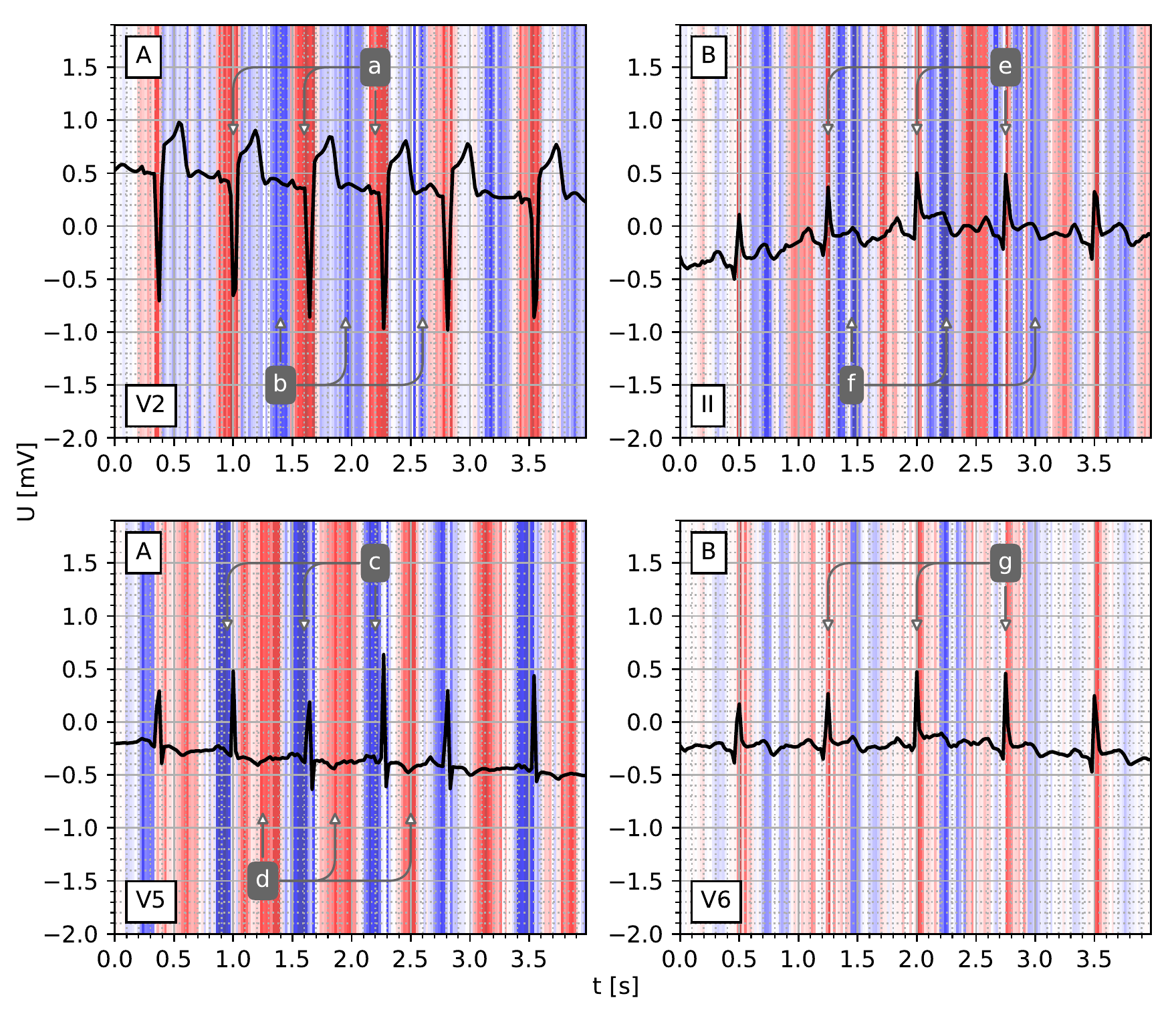}

    \caption{Examplary interpretability analysis for two ECGs with myocardial
    infarction. The attribution score is color-coded in the background. Red
    (blue) areas influenced the neural network towards (against) a correct
    classification as myocardial infarction. See the accompanying text for details.}

    \label{fig:interpret_small}
\end{figure}

Fig.~\ref{fig:interpret_small} shows examples of the interpretability analysis
of selected channels of two myocardial infarction ECGs that were
correctly classified. As in the clinical context ECGs are always considered in
the context of the full set of twelve channels (if available), the complete set
of channels is shown in Fig.~\ref{fig:interpret_large} in the appendix. Take
note that the attributions show a high consistency over beats of one ECG, even
if a significant baseline shift is present. There is also a reasonable
consistency with regard to similar ECG features exhibited by other patients
which are not shown here.

ECG A is taken from a 74-year-old male patient one day after the infarction took
place. A coronary angiography performed later confirmed an anterior myocardial
infarction. ECG B is taken from a 68-year-old male patient one day after the
infarction which was later confirmed to be in inferior localization. ECGs A and
B are listed as s0021are and s0225lre in the PTB dataset.

Signs for ischemia and infarction are numerous and of variable specificity \cite{pmid28611879}.
Highlighted areas coincide with established ECG signs of myocardial infarction.
These are typically found between and including the QRS complex and the T wave, as this
is when the contraction and consecutive repolarization of the ventricles take place.
ST-segment-elevation (STE) is the most important finding in myocardial infarction
ECGs and diagnostic criterion for ST-elevation myocardial infarction (STEMI) \cite{pmid25689940}.
This sign (though not formally significant in every case) and corresponding high
attributions can be found in both example ECGs at the positions marked
\textbf{a}, \textbf{e}, and \textbf{g}. At the same time in another channel (position \textbf{c}) there is no STE
and the attribution is consequently inverted.
The attribution at position \textbf{a} also coincides with pathological Q waves,
which also occur in some infarction ECGs.
T wave inversion, another common sign for infarction, can be found at position \textbf{d}.
Some other attributions of the model are less conclusive. Although attributions at positions \textbf{b} and \textbf{f} fall
in the T and/or U waves, that is regions that are relevant for detection of infarction,
it is unclear why they influenced the decision against infarction.

Note that the highlighted areas do not necessarily align perfectly with what clinicians
would identify as important. For example for a convolutional neural network to detect an
ST-elevation, it must use and compare information from before and after the QRS complex,
which most likely results in high attributions to the QRS complex itself and its
immediate surrounding rather than to the elevated ST-segment.

Comparing the overall visual impression of the attributions across all channels (see Fig.~\ref{fig:interpret_large}),
the model seems to attribute more importance to the Wilson leads in ECG A (anterior infarction) and more
importance to the limb leads in ECG B (inferior infarction). This is also where clinicians would expect to find
signs of infarction in these cases.

Attributions are inherently model-dependent and as a matter of fact the corresponding attributions
show quantitative and in some cases even qualitative differences. However, across different folds and different
random initializations the attribution corresponding to the STE was always correctly and prominently identified.
This is a very encouraging sign for future classification studies on ECG data based on convolutional methods,
in particular in combination with attribution methods. A future study could put
the qualitative finding presented in this section on a quantitative basis. This would require
a segmentation of the data, possibly using another model trained on an annotated
dataset as no annotations are available for the PTB dataset, and statistically evaluating
attribution scores in conjunction with this information. In this context it would be
interesting to see if different classification patterns arise across different models or
if they can at least be enforced as in \cite{DBLP:journals/corr/RossHD17}.

\section{Summary and Conclusions}

In this work, we put forward a fully convolutional neural network for myocardial
infarction detection evaluated on the PTB dataset. The proposed
architecture outperforms the current state-of-the-art approaches on this dataset and
reaches a similar level of performance as human cardiologists for this task. We investigate
the classification performance on subdiagnoses and identify two clinically
well-motivated subdiagnosis classes that can be separated very well by our algorithm.
We focus on the clinically most relevant case of 12-lead data and stress the
importance of a careful data selection and cross-validation procedure.

Moreover, we present a first exploratory study of the application of
interpretability methods in this domain, which is a key requirement for
applications in the medical field. These methods can not only help to gain
an understanding and thereby build trust in the network's decision process
but could also lead to a data-driven identification of important markers for
certain classification decisions in ECG data that might even prove useful
to human experts. Here we identified common cardiologists' decision rules
in the network's attribution maps and outlined prospects for future studies in this
direction.

Both such an analysis of attribution maps and further improvements of
the classification performance would have to rely on considerably larger
databases such as \cite{couderc2010unique} for quantitative precision. This would
also allow an extension to further subdiagnoses and other cardiac conditions such
as other confounding and non-exclusive diagnoses or irregular heart rhythms.

\appendix
\section{Network architectures}
\label{app:architectures}
All models were implemented in TensorFlow \cite{Abadi2016tensorflow}. As only preprocessing step
we apply input normalization by applying batch normalization \cite{2015arXiv150203167I} to all input channels. In all cases we minimize
crossentropy loss using the Adam optimizer \cite{2014arXiv1412.6980K} with learning rate 0.001.

A uniformly sampled ECG signal can be represented as a two-dimensional tensor ($\#$ sampling points $\times$ input channels), as opposed to 
image data that is conventionally represented as a three-dimensional tensor (height $\times$ width $\times$ input channels), to which 
for example conventional CNN building blocks like one-dimensional convolutional or pooling layer (operating on a single axis rather than two axes 
in the image case) can be applied straightforwardly. This approach is predominantly used in the literature, see e.g.\ \cite{Zheng2014,ACHARYA201762,rajpurkar_cardiologist-level_2017}.

\subsection{Convolutional architectures}
\begin{figure}[ht]
    \centering
    \includegraphics[width=\columnwidth]{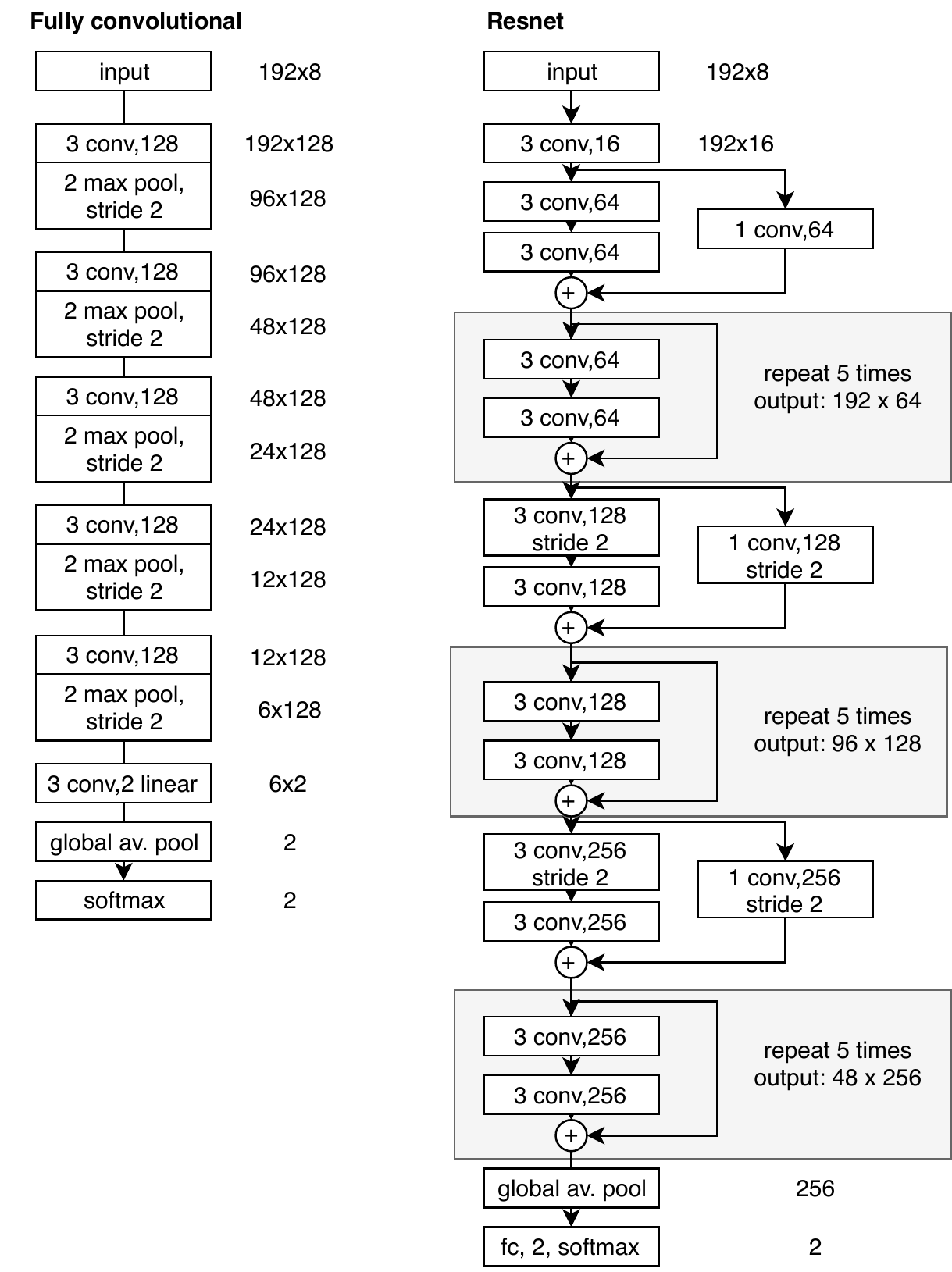}
	\caption{Convolutional architectures: fully convolutional (left) and resnet (right).}
    \label{fig:architectures}

\end{figure}

We generally use ELU \cite{DBLP:journals/corr/ClevertUH15} as activation function both for
convolutional as well as fully connected layers without using batch normalization \cite{2015arXiv150203167I},
which was reported to lead to a slight performance increase compared to the standard ReLU activation
with batch normalization \cite{2016arXiv160602228M}. In the architectures with fully connected layers we apply dropout \cite{srivastava_dropout:_2014}
at a rate of 0.5 to improve the generalization capability of the model. We initialize weights according to \cite{he2015delving}. Note that in contrast to the case of two-dimensional data
a max pooling operation only reduces the number of couplings by a factor of 2 rather than 4, which is then fully compensated
by the conventional increase of filter dimensions by 2 in the next convolutional layer. To achieve a gradual reduction of couplings we therefore
keep the number of filters constant across convolutional layers. We study the following convolutional architectures that are also
depicted in Fig.~\ref{fig:architectures}:
\begin{enumerate}
\item A fully convolutional architecture \cite{long_fully_2015} with a final global average pooling layer
\item A resnet-inspired \cite{2015arXiv151203385H,2016arXiv160305027H,2016arXiv160507146Z} architecture with skip-connections
\end{enumerate}
We investigate the impact of including frequency information obtained via a Fast Fourier Transformation with $N_\text{FFT}=2^{\ceil{ \log_2 (d)}}$, where $d=192$ denotes the sequence length after rescaling.
The $N_\text{FFT}/2+1$ independent components are used as frequency-domain input data with otherwise unchanged network architectures.

\subsection{Recurrent architectures}
\label{app:recurrent}
As alternative architecture we investigate recurrent neural networks, namely LSTM \cite{hochreiter_long_1997} cells.
We investigated stacked LSTM architectures but found no significant gain in performance. However, even for
a single RNN cell, in our case with 256 hidden units, different training methods are feasible:
\begin{enumerate}
\item In the first variant we feed the last LSTM output into a fully connected softmax layer.
\item In the second variant we additionally apply a time-distributed fully connected layer, i.e.\ a fully connected
layer with shared weights for every timestep, and train the network to predict the next element in a
time series prediction task jointly with the classification task. Here we
adjusted both loss functions to reach similar values. Similar to
\cite{Husken2003} we investigate in this way if the time series predication task
improves the classification accuracy.
\end{enumerate}
During RNN training we apply gradient clipping.

\section*{Acknowledgment}
The authors thank M.~Gr\"unewald and K.-R.~M\"uller for discussions and E.~Dolman for comments on the manuscript.

\bibliographystyle{IEEEtran}
\bibliography{bibfile}

\begin{landscape}
    \begin{figure}
    	\centering
    	\includegraphics[width=\columnwidth]{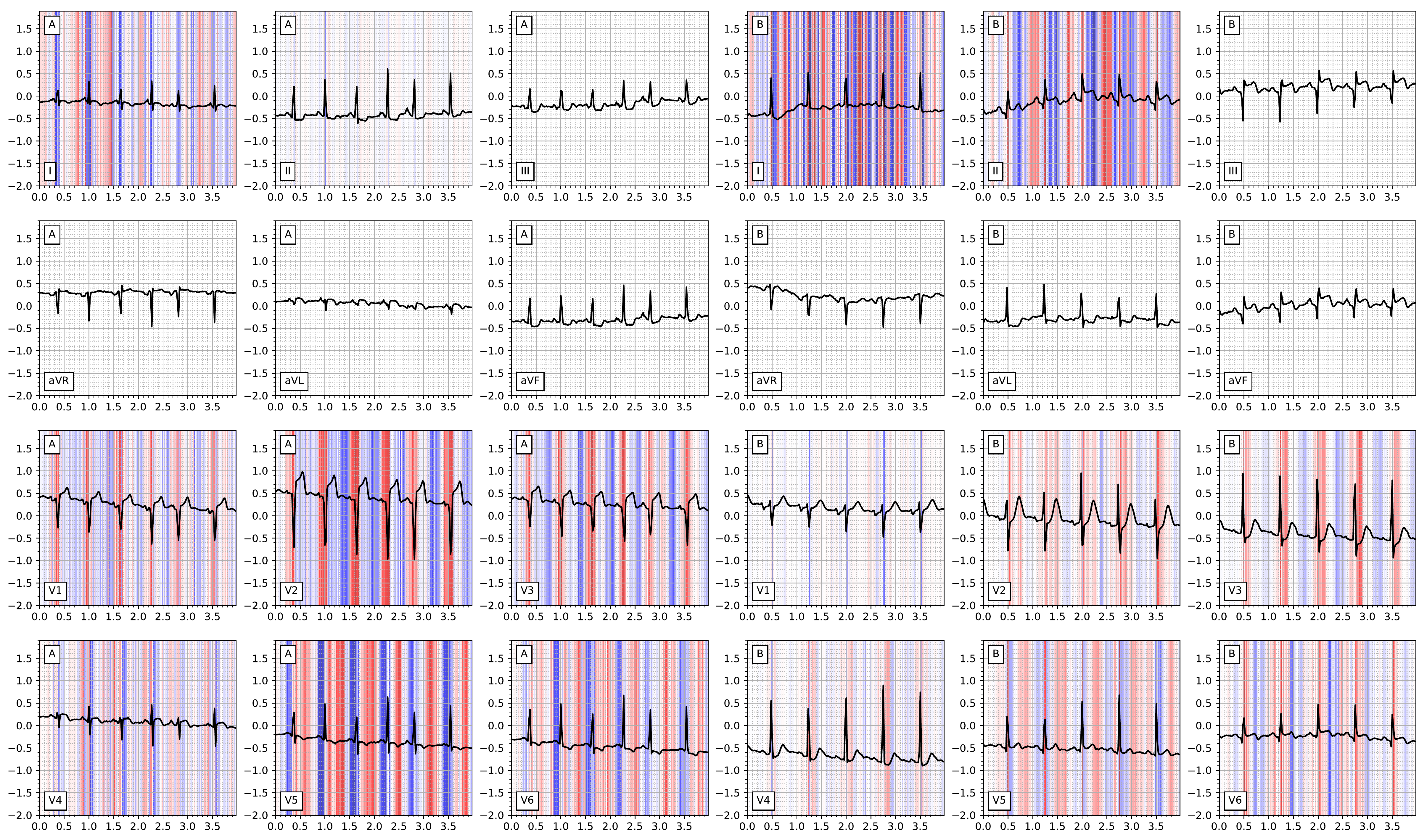}

        \caption{Examplary interpretability analysis repeated here with the full
        set of channels. Red (blue) areas influenced the neural network towards (against) a correct
        classification as myocardial infarction. Note that the neural network did only use a
        non-redundant set of 8 instead of the full set of 12 channels.
        Nevertheless, unused channels are shown as well without color-coding in
        the background. See Sec.~\ref{sec:results_interpretability} for
        details.}

        \label{fig:interpret_large}
    \end{figure}
\end{landscape}
\end{document}